\documentclass[twocolumn,showpacs,preprintnumbers,amsmath,amssymb,prb]{revtex4}

\usepackage{graphicx}
\usepackage{dcolumn}
\usepackage{bm}
\usepackage{textcomp}
\usepackage{ulem}
\usepackage{multirow}
\usepackage[latin1]{inputenc}
\usepackage{enumerate}

\usepackage{amstext}

\begin{document}


\title{Combinatorial model for the ferroelectric domain-network formation in hexagonal manganites}

\author{Bruno Mettout}
\author{Pierre Tolédano}
 \affiliation{Laboratoire de Physique des Systemes Complexes, Université de Picardie, 80000 Amiens, France}
\author{Martin Lilienblum}
\author{Manfred Fiebig}
 \affiliation{Department of Materials, ETH Zurich, Wolfgang-Pauli-Strasse 10, 8093 Zurich, Switzerland}

\date{\today}

\begin{abstract}
As the result of the high-temperature improper ferroelectric transition occurring in YMnO$_3$ and
other hexagonal manganites a striking domain pattern was observed: an emergence of six domains of
alternating polarization from one point in a vortex-like pattern. We derive the formation and
distribution of the domain-vortex network from a combinatorial analysis based on order-parameter
symmetry and lattice geometry. The analysis leads to stable vortex-like configurations of the six
domain states in the basal plane perpendicular to the polarization and to fragmented vortices in
planes parallel to the polarization. The predictions are debated in the light of existing
experimental and theoretical work on the vortex domain state. The relation of the
symmetry-determined geometrical defects to the topological defects proposed to describe the
behavior of YMnO$_3$ near the Curie temperature is discussed.
\end{abstract}


\maketitle


\section{Introduction: Ferroelectric vortices in hexagonal manganites}

After there has been some debate concerning the thermodynamic path occurring at the transition
between the paraelectric $P6_3/mmc$ (Z=2) and ferroelectric $P6_3cm$ (Z=6) phases in YMnO$_3$
recent experimental and theoretical investigations agree in that the cell-tripled improper
ferroelectric structure arises directly at the Curie temperature $T_\text{C}$ of about
1260~K.\cite{Ismailzade65, Lukaszewicz74, Abrahams01, Katsufuji02, Lonkai04, VanAken04, Nenert07,
Fennie05, Gibbs11} A confirmation of this direct path is the observation of a ferroelectric domain
pattern below $T_\text{C}$ consistent with the two-dimensional-order-parameter symmetry associated
with the $K_3$ zone-boundary phonon mode at $\vec{k}=(\frac{1}{3}, \frac{1}{3}, 0)$.\cite{Chae10,
Choi10, Jungk10, Lilienblum11, Chae12, Zhang13, Meier13} The corresponding symmetry-breaking
mechanism gives rise to a total of six domain states. These combine three antiphase domain states
resulting from the loss of the $(a,0,0)$, $(a,a,0)$ and $(0,a,0)$ paraelectric translations with
two $180^{\circ}$ domain states representing opposite spontaneous polarization along $(0,0,c)$.

\begin{figure}
    \centering
        \includegraphics[width=\columnwidth,clip]{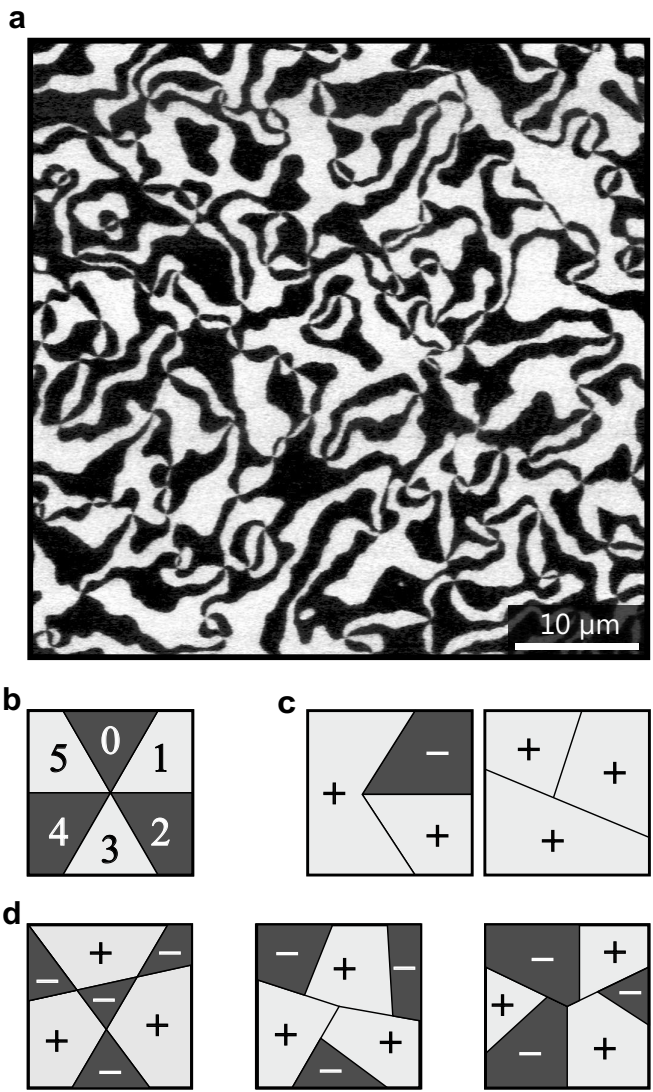}
\caption{(a) Domain pattern observed by piezoresponse force microscopy in the basal $(a,b)$ plane
of hexagonal YMnO$_3$. The image was taken from Ref.~\protect\onlinecite{Jungk10}. (b)-(d)
Schematic domain configurations that are in principle possible and allowed by the order-parameter
symmetry: (b) six-branch vortex-like pattern; (c) three-branch domain patterns; (d) combinations
into multi-centred fragmented patterns.}
    \label{fig:Figure1}
\end{figure}

The most surprising feature of the domain pattern observed in the ferroelectric phase of YMnO$_3$
[Fig.~\ref{fig:Figure1}(a)] is that the six domain states form an exclusive vortex-like pattern
with six domains of alternating polarization emerging from one point (Fig.~\ref{fig:Figure1}(b)).
In contrast, $N$-branched arrangements of domains with $N\neq 6$ [Fig.~\ref{fig:Figure1}(c)] or
their combination into fragmented domain configurations [Fig.~\ref{fig:Figure1}(d)] are absent.
The absence of three-branched domains is particularly puzzling since from a geometrical point of
view these states are structurally stable: A shift of a domain wall may shift the point of
intersection but this would not break up the three-branch configuration as such. By contrast, a
single-center six-branched interlocked domain state is geometrically less stable as it may split
into fragmented domains under any small deformation [Fig.~\ref{fig:Figure1}(d)].

The vortex-like six-branch domain pattern in the hexagonal manganites has been investigated in a
variety of approaches so far. Chae et al.\ analyzed the network of ferroelectric domains by graph
theory\cite{Chae10} and Monte-Carlo simulations.\cite{Chae12} Their work is focused on the
\textit{description} of the experimentally observed domain pattern which is based on the
distribution of a single type of vortex and antivortex. A \textit{derivation} of the structure of
the vortices and the resulting domain network was not intended. Artyukhin et al.\ apply Landau
theory\cite{Artyukhin13} for deriving the energetically most beneficial arrangement of domain
states around the vortex and find that this is a configuration with six states of alternating
polarization. Kumagai et al.\ investigate the structure and energy of the walls between domain
states by density functional theory\cite{Kumagai13} and confirm the arrangement of states with
alternating polarization. All these investigations, however, leave the question of the structure
of the domain vortex point on the length scale of the unit cell open. This issue was investigated
by transmission electron microscopy with controversial results. Yu et al.\ report that on the
length scale of a unit cell the domains avoid one another so that there is no actual point in
which all the domain states meet.\cite{Yu13} In contrast, Zhang et al.\ observe a meeting of all
six domain states within a diameter of four unit cells.\cite{Zhang13}

In this report we derive the structure of the domain vortex state in the hexagonal manganites from
a combinatorial analysis based on symmetry and geometry. We find that \textit{three} classes of
six-branch domain (anti-) vortices are compatible with the site symmetry of the hexagonal
manganites, even though only one of these classes is realized. Vortices in the basal $(a,b)$ plane
originate in the $\overline{6}m2$ high-symmetry lattice point of the paraelectric unit cell. In
contrast, vortices in planes parallel to the $c$ axis are expected to fragment which is supported by
recent electron microscopy results. Aspects of topology of the domain vortices and the relation to
other types of vortex domain structures are discussed.

\section{Combinatorial model for domain vortices}

\subsection{Symmetry}

\begin{figure}
    \centering
        \includegraphics[width=\columnwidth,clip]{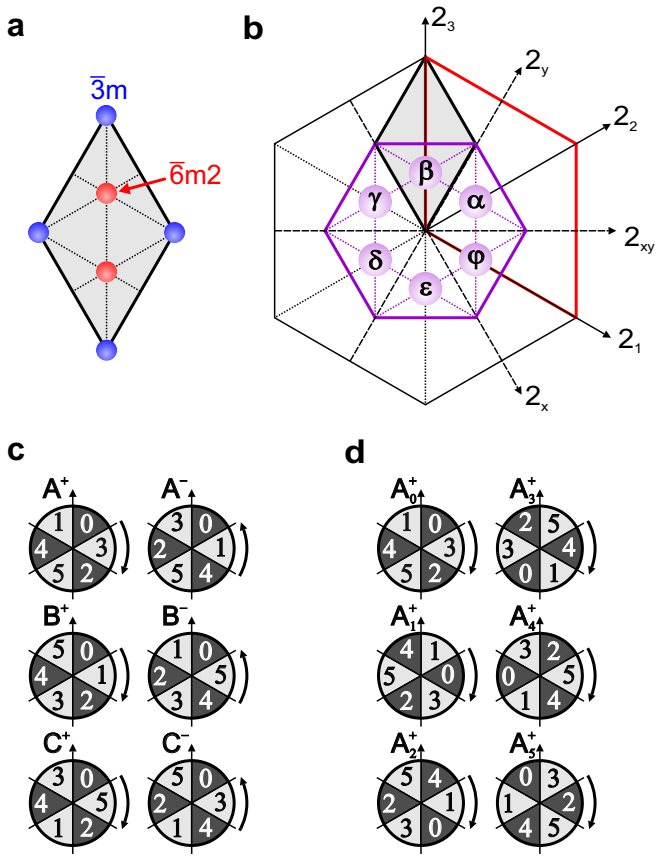}
\caption{(a) High-symmetry sites in the basal $(a, b)$ plane of the paraelectric $P6_3/mmc$
unit-cell of the hexagonal manganites. Red ions: O$^{2-}$, blue ions: Y$^{3+}$. (b) Location of
the six sites $\alpha$ to $\varphi$ with $\overline{6}m2$ symmetry in relation to the tripled
ferroelectric unit-cell. This permits the formation of a domain vortex pattern which has the
chiral point symmetry 32, a subgroup of $\overline{6}m2$. (c) The three classes of
vortex-antivortex domain configurations $A^\pm ,B^\pm ,C^\pm$. (d) The six domain configurations
forming the $A^+$ class.}
    \label{fig:Figure2}
\end{figure}

The exclusive stabilization of the six-branched domain-state vortex shown in
Fig.~\ref{fig:Figure1}(b) in favor of vortices involving a different number of domain states
[Fig.~\ref{fig:Figure1}(c)] or fragmentation [Fig.~\ref{fig:Figure1}(d)] points to symmetry
criteria that render the $N\neq 6$ and fragmented states unstable. Finding this higher symmetry
requires to determine at which points of the paraelectric YMnO$_3$ lattice a center point of the
domain vortex can be created. The criterion is that the site symmetry of the YMnO$_3$ lattice at
the position of this center point includes the symmetry of the domain vortex pattern around this
point. Only crystallographic sites in the basal $(a,b)$ plane of the paraelectric space-group
$P6_3/mmc$ [Fig.~\ref{fig:Figure2}(a)] with symmetries $\overline{3}m$ $(D_{3d})$ (Wyckoff
a-position, Y$^{3+}$ site) or $\overline{6}m2$ $(D_{3h})$ (Wyckoff c- and d-positions, Mn$^{3+}$
site) verify this property. Applying the symmetry operations of the $P6_3/mmc$ lattice to the
domain vortex pattern shows that \textit{its highest possible symmetry belongs to the} $32$
$(D_3)$ \textit{chiral point-group} which is a subgroup of $\overline{6}m2$.
Figure~\ref{fig:Figure2}(b) shows the six $\overline{6}m2$ sites, denoted $\alpha, \beta, \gamma,
\delta, \epsilon, \varphi$, around which a domain vortex pattern can form within the YMnO$_3$
lattice. They are located at the center of triangles forming the Wigner-Seitz cell of the tripled
ferroelectric $P6_3cm$ unit cell.

The point symmetry 32 of the domain vortex selects 36 configurations among 6! = 720 possible
sequences of domain states around the vortex center. The 684 remaining configurations have a lower
symmetry and are therefore unstable and prone to fragmentation. The 36 allowed configurations are
divided into three classes of vortex-antivortex configurations denoted $A^\pm ,B^\pm ,C^\pm$
[Fig.~\ref{fig:Figure2}(c)]. They differ by the order of the domain sequence: $A^+$ (032541),
$B^+$ (012345),  $C^+$ (052143) and the reversed sequences for $A^- ,B^- ,C^-$. Each vortex or
antivortex class can exist in six different configurations depending on the position of the
domains with respect to an arbitrary fixed direction. Figure~\ref{fig:Figure2}(d) shows the six
configurations of the $A^+$ class which are denoted $A^+_0, A^+_1, A^+_2, A^+_3, A^+_4, A^+_5$
depending on the location of the domains 0,1,2,3,4,5 with respect to a fixed in-plane axis, e.g.\
the two-fold rotation $2_3$ of the 32 group. Each of the six sites ($\alpha$ to $\varphi$) is
compatible with only six vortex configurations as listed in Table \ref{tab:table1}: the $(\alpha,
\gamma, \epsilon)$ sites are associated with the ($A^+, B^+, C^+$) vortices whereas ($\beta,
\delta, \phi$) sites correspond to the ($A^-, B^-, C^-$) vortices.

\begin{table}[t!]
    \centering
        \begin{tabular}{rr|rrrrrr|}
            $\alpha$ & & $A^+_0$ & $B^+_5$ & $C^+_1$ & $B^+_2$ & $A^+_3$ & $C^+_4$ \\
            $\beta$ & & $B^-_0$ & $A^-_5$ & $C^-_1$ & $A^-_2$ & $B^-_3$ & $C^-_4$ \\
            $\gamma$ & & $B^+_0$ & $C^+_5$ & $A^+_1$ & $C^+_2$ & $B^+_3$ & $A^+_4$ \\
            $\delta$ & & $A^-_0$ & $C^-_5$ & $B^-_1$ & $C^-_2$ & $A^-_3$ & $B^-_4$ \\
          $\epsilon$ & & $C^+_0$ & $A^+_5$ & $B^+_1$ & $A^+_2$ & $C^+_3$ & $A^+_4$ \\
            $\varphi$ & & $C^-_0$ & $B^-_5$ & $A^-_1$ & $B^-_2$ & $C^-_3$ & $A^-_4$ \\
        \end{tabular}
\caption{Distribution of vortex configurations among the six lattice sites $\alpha$ to $\varphi$
shown in Fig.~\ref{fig:Figure2}(b).}
    \label{tab:table1}
\end{table}

\subsection{Connections between two vortices}

\begin{figure*}
    \centering
        \includegraphics[width=2\columnwidth,clip]{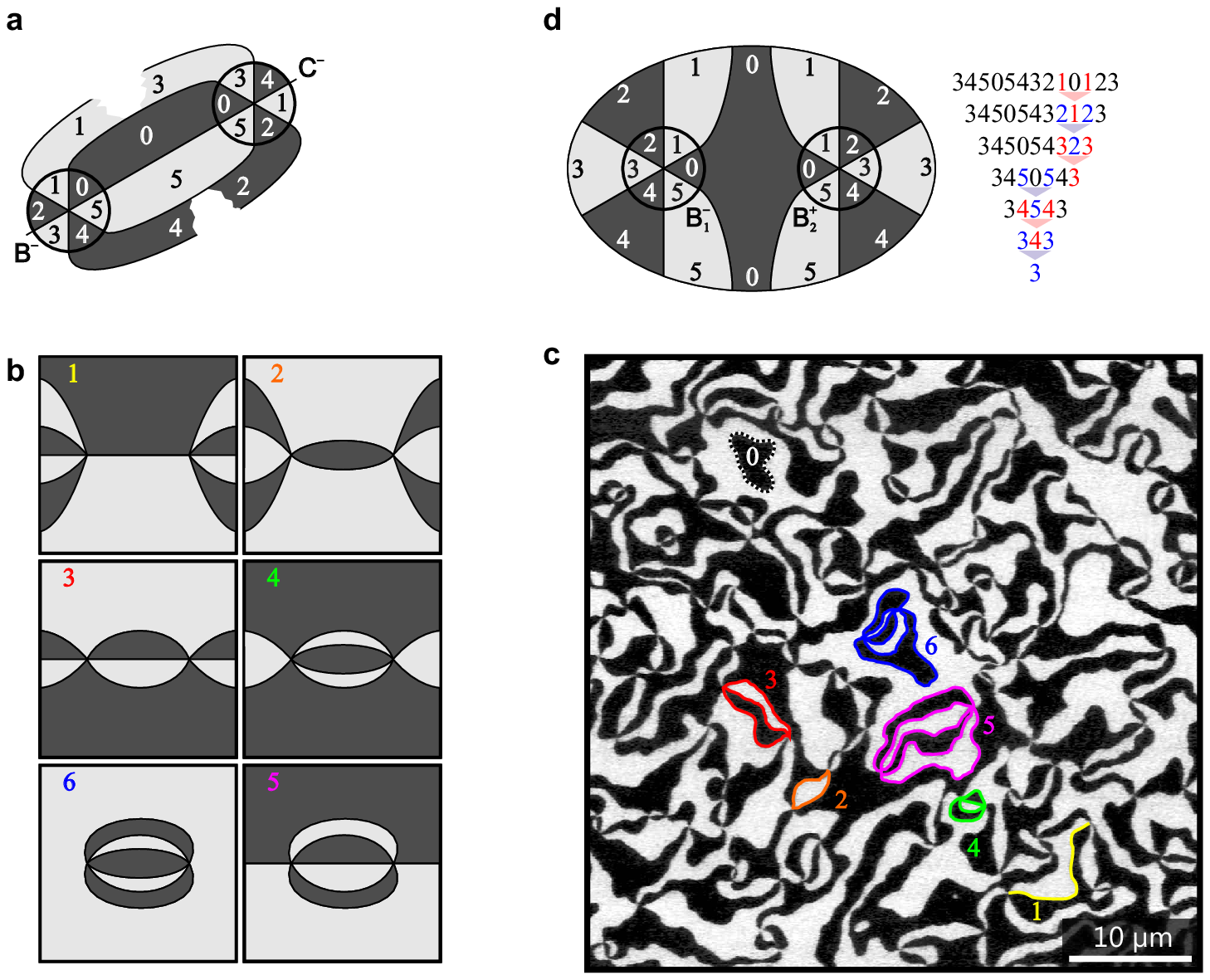}
    \caption{(a) Illustration of the impossibility of joining two
vortices belonging to different classes by more than one curve. (b) Possible connections by 1 to 6
domain walls (labels) between a vortex and its antivortex, and (c) their occurrence in the
YMnO$_3$ domain network. (d) Contraction mechanism reducing the number of connecting curves for
the $B^-_1 - B^+_2$ pair.}
    \label{fig:Figure3}
\end{figure*}

From the preceding considerations one can deduce the rules which determine the connectivity of a
domain network with A, B, and C-type vortices: \textit{Two vortices can be connected by one to six
curves separating two domains as shown in Fig.~\ref{fig:Figure3}. Rule 1: A connection involving
more than one curve occurs between a vortex and its antivortex; Rule 2: A connection by a single
curve can occur either between a vortex and its antivortex or between two vortices or two
antivortices of a univocally determined couple of different classes.} Rule 1 is deduced from the
property that a vortex is fully determined by an ordered sequence of three numbers. As shown in
Fig.~\ref{fig:Figure2}(a) the sequences (032), (123) and (052) designate the vortices $A^+, B^+$
and $C^+$, respectively. Turning in the opposite sense around a vortex, the same sequence of
numbers are exclusive to the corresponding antivortices $A^-, B^-$ and $C^-$, respectively.
Fig.~\ref{fig:Figure3}(a) illustrates the impossibility of connecting two vortices belonging to
different classes by more than one curve. Rule 2 stems from the property that an ordered sequence
of two numbers always typifies two vortex classes: For example, the (30) sequence characterizes
the $C^+$ and $A^-$  vortices whereas the reversed sequence (03) is associated with $C^-$ and
$A^+$, so that  $C^+$ and $A^+$ can be connected by a single curve, while an $A^+$ vortex cannot
be paired with a $B^+$ or $A^+$ vortex. Table \ref{tab:table2} lists the couples of vortices which
can be connected by a single curve.

\begin{table}[t!]
    \centering
        \begin{tabular}{cc|cccccc}
\textbf{} & & \textbf{0} & \textbf{1} & \textbf{2} & \textbf{3} & \textbf{4} & \textbf{5} \\
\hline
\textbf{0} & & & $A^-B^+$ &  & $A^+C^-$ &  & $B^-C^+$ \\
\textbf{1} & & $A^+B^-$ & $$ & $B^+C^-$ & $$ & $A^-C^+$ & $$ \\
\textbf{2} & & $$ & $B^-C^+$ & $$ & $A^-B^+$ & $$ & $A^+C^-$ \\
\textbf{3} & & $A^-C^+$ & $$ & $A^+B^-$ & $$ & $B^+C^-$ & $$ \\
\textbf{4} & & $$ & $A^+C^-$ & $$ & $B^-C^+$ & $$ & $B^+A^-$ \\
\textbf{5} & & $B^+C^-$ & $$ & $A^-C^+$ & $$ & $A^+B^-$ & $$ \\
        \end{tabular}
    \caption{Pair of neighboring domain states (bold numbers) and the associated couple of vortex classes where
    they occur. For example, the sequence 43 of domain states occurs in the classes $B^-$ and
    $C^+$.}
    \label{tab:table2}
\end{table}

Figure~\ref{fig:Figure3}(b) summarizes the different types of connections observed between domain
vortices in the network shown in Fig.~\ref{fig:Figure3}(c). To begin with, a single
vortex-antivortex pair connected by six curves is found, forming an ``island'' (label ``6'')
surrounded by a single domain. The corresponding pair is topologically trivial since a loop
surrounding it runs within a single domain, justifying the terms vortex/antivortex. Two pairs
forming ``bubbles'' (label ``5'') surrounded by domains of opposed polarities are observed in
Fig.~\ref{fig:Figure3}(c), each pair being interconnected by five curves. The number of
vortex-antivortex pairs increases with the decreasing number of curves connecting them, i.e.
seven, eleven, and seventeen pairs are found in Fig.~\ref{fig:Figure3}(c) for two vortices
connected by, respectively, four, three, and one (labels accordingly) curves.

This can be illustrated by the following \textit{contraction mechanism} which is required for
increasing the number of connections between a vortex and its antivortex. The mechanism shown in
Fig.~\ref{fig:Figure3}(d) consists of the removal of an interstitial domain between two identical
domains in the loop surrounding a pair of domain vortices. By the removal the two identical
domains are merged into a single one, and the connectivity of the pair of domain vortices is
increased by one. Thus, step by step, the connectivity is increased by decreasing the number of
domains crossed by the loop surrounding the domain-vortex pair until, in the ultimate case, an
island-like pair of a vortex and an antivortex with a connectivity of six is obtained. The
surrounding domain is determined by the order obeyed during the sequence of successive
contractions. If the contraction does not proceed as shown in Fig.~\ref{fig:Figure3}(d), because
of interaction with other nearby vortices or if the contraction involves vortices from different
classes, the contraction mechanism leads to a smaller number of connecting curves between the
vortices forming a pair.

\begin{figure*}
    \centering
        \includegraphics[width=2\columnwidth,clip]{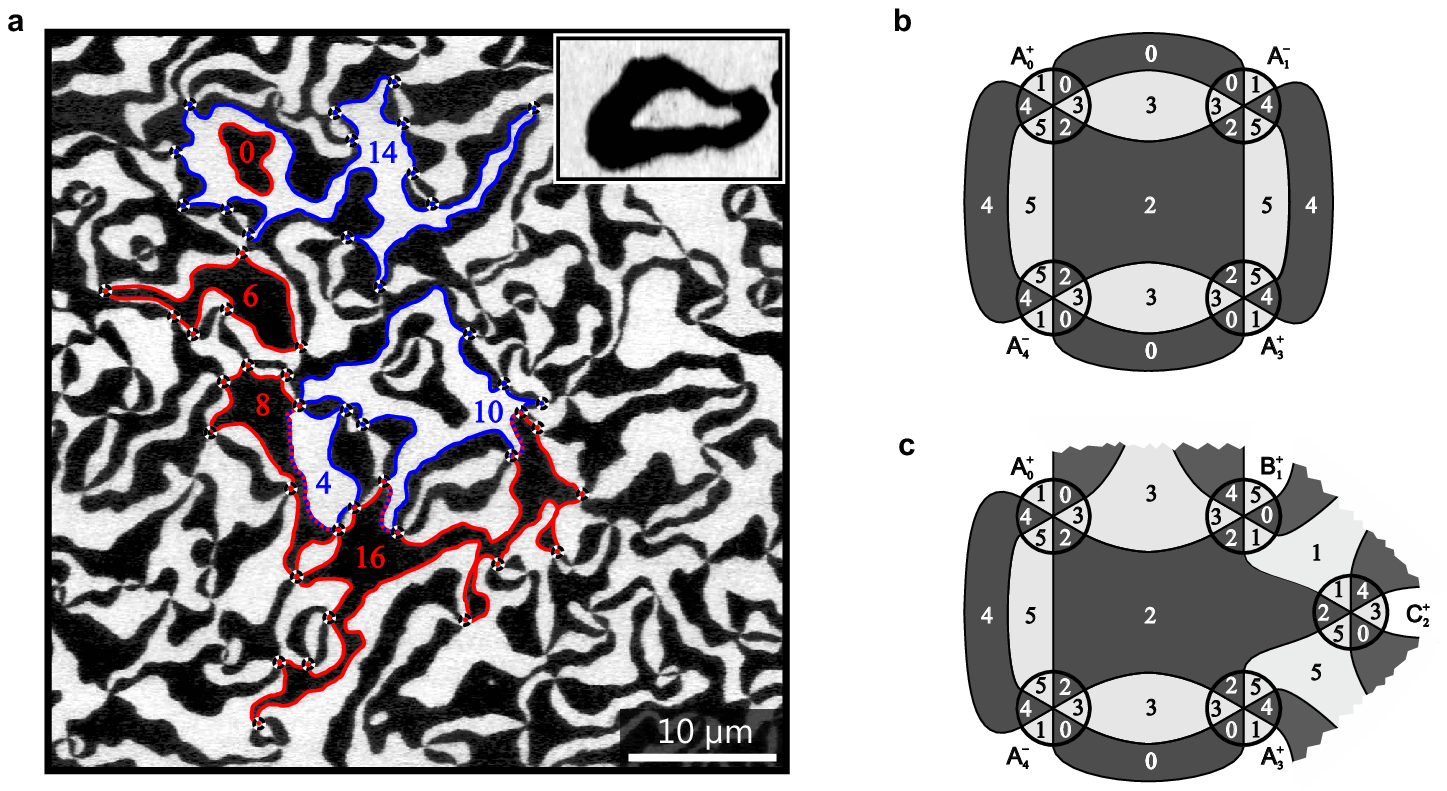}
    \caption{(a) Examples of $n$-gon vortex configurations with $n = 0, 2, 4, 6, 8, 10, 14, 16$
observed in YMnO$_3$. Annular domains in the form of a 0-gon embedded in another 0-gon as shown in
the inset and discussed in Ref.~\protect\onlinecite{Li13} are an expected variation of this
arrangement. (b) Sketch of an even-gon configuration involving vortex-antivortex pairs belonging
to the same class. (c) Sketch of an odd-gon configuration involving vortices from different
classes (not observed).}
    \label{fig:Figure4}
\end{figure*}

\subsection{Type of vortex networks}

The preceding discussion was restricted to the case of two vortices being connected in forming a
domain, i.e., a 2-gon configuration. Figure~\ref{fig:Figure4}(a) shows that $n$-gon vortex
configurations with $n$ even and ranging from 0 to at least 16 are observed in YMnO$_3$. The
restriction to even-gons, which laid the basis for the graph-theoretical description by Chae et
al.\cite{Chae10}, points to a pattern consisting of a single class of vortex and antivortex. These
can only occur in pairs since identical vortices (like $A^+ - A^+$), as they would be present in
the associated odd-gons, cannot be connected, see Rules 1 and 2. In principle, even-gons and
odd-gons composed of different types of \mbox{(anti-)vortices} like the one in
Fig.~\ref{fig:Figure4}(c) are allowed by rule 2 but they are never observed. Thus, applying the
connectivity rules to the observed domain pattern leads to the conclusion that \textit{only one
type of vortex-antivortex pair is present in YMnO$_3$}. Furthermore, applying the paraelectric
symmetry operations to the three classes of vortices shows that subsets of vortices within the
$A^\pm$ and $C^\pm$ classes transform into one another under these symmetry operations. Hence,
they are energetically equivalent and should therefore both be present or absent in the domain
pattern. By contrast, $B^\pm$ vortices always transform into $B^\pm$ vortices. Therefore, one can
\textit{unambiguously identify the type of the exclusively occurring class of vortices as B}.
Table \ref{tab:table1} provides the type of $B^\pm$ vortex associated with each $\overline{6}m2$
site in Fig.~\ref{fig:Figure2}(d).

Two aspects remain to be understood, namely (i) the process leading to the formation of vortices
and vortex pairs; (ii) the restriction to a solitary type of vortex and antivortex.

\begin{figure}
    \centering
        \includegraphics[width=\columnwidth,clip]{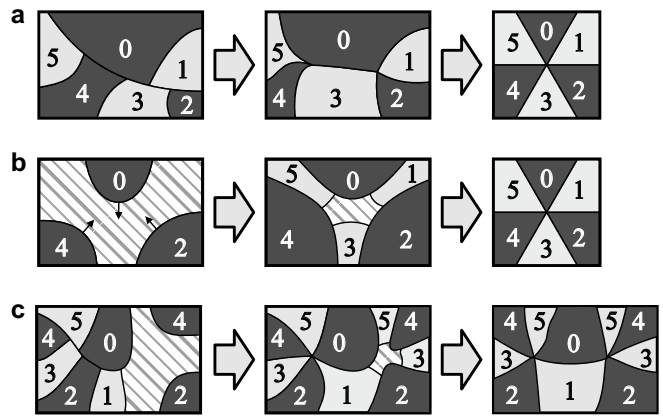}
    \caption{(a) Defragmentation process and (b) insertion of interstitial
domains leading to a six-branch domain-vortex. (c) Combined mechanisms contributing to the
formation of a vortex-antivortex pair.}
    \label{fig:Figure5}
\end{figure}

Regarding (i), we assume, supported by theory,\cite{Kumagai13} that interfaces between
odd-numbered or even-numbered domains are so energetically unfavorable that they are strictly
avoided. Therefore, when two domains of same parity, e.g. 0 and 2, get closer they form an
intermediate region of opposed parity (012) with the creation of an additional domain-wall. This
explains the scarcity of vortex-antivortex pairs connected by a single curve. The interstitial
domain created between even or odd domains within the $B^\pm$ pairs is pre-determined as shown in
Fig.~\ref{fig:Figure2}(c): the domain state 1 is formed between 0 and 2, domain state 4 separates
3 and 5 etc. Figures~\ref{fig:Figure5}(a) and \ref{fig:Figure5}(b) illustrate the according
defragmentation process and the creation of interstitial domains leading to the formation of a
$B^+$ vortex. Figure~\ref{fig:Figure5}(c) shows how these mechanisms can combine to form a
vortex-antivortex pair.

Regarding (ii), we assume that interfaces between domain states as they would occur in A- or
C-type vortices are energetically unfavorable in comparison to the interfaces in B-type vortices.
Therefore, when two such domain states (such as 0 and 3) get closer, they form intermediate domain
states (like 0123) with additional walls of reduced energy. This is supported by Landau and
density-functional theory which show that the B-type vortices indeed represent the arrangement
with the lowest energy between neighboring domains.\cite{Artyukhin13, Kumagai13} Walls between
different antiphase domains of different polarization possess the lowest energy. Walls between
different antiphase domains of the same polarization are less preferred, and walls between the
same type of antiphase domain but opposite polarization are least favorable.\cite{Kumagai13} These
results lead to the B-type vortices if we associate the sequence 0, 1, 2, 3, 4, 5 of domains to
the physical domain states $\alpha^+$, $\beta^-$, $\gamma^+$, $\alpha^-$, $\beta^+$, $\gamma^-$.

Note that according to our combinatorial derivation other domain patterns than in the hexagonal
manganites may occur in isostructural materials with different energies of the domain states and
the walls separating them. An investigation of the hexagonal ferrites might prove interesting in
this respect.\cite{Wang13} In addition, the preference of a symmetric non-fragmented vortex over a
lower-symmetry fragmented one is not necessarily universal. A modification of the thermodynamic
parameters might interchange the stability of the two types of configurations. Thus, varying the
temperature can, in principle, lead to a \textit{fragmentation transition}. However, this has not
yet been observed. Furthermore, because of the perturbation exerted by the polarization field near
the vortex core a slight defragmentation on the length scale below the resolution limit of force
microscopy experiments may occur after all. This will be investigated further in the following
section.

\begin{figure}
    \centering
    \includegraphics[width=\columnwidth,clip]{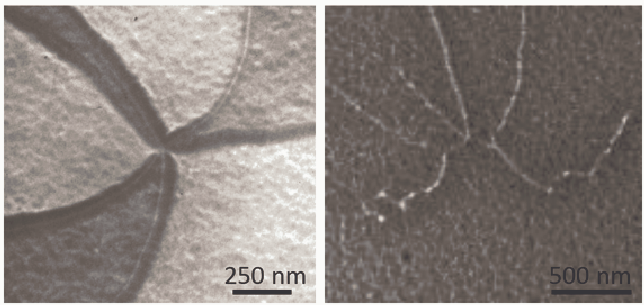}
    \caption{Electron microscopy images of ErMnO$_3$ on faces (a) perpendicular to,
and (b) including the direction of the ferroelectric polarization (adapted from
Ref.~\protect\onlinecite{Meier14}).}
    \label{fig:Figure6}
\end{figure}

\section{Experimental investigation of the vortex center}

Figure~\ref{fig:Figure6} shows low-energy electron microcopy data adapted from
Ref.~\onlinecite{Meier14}. The images show two faces of an ErMnO$_3$ sample oriented,
respectively, perpendicular (basal $(a,b)$ plane) and parallel to the direction of the spontaneous
polarization. While the domain vortex on the $(a,b)$ plane does not show any fragmentation within
the resolution limit of about 5~nm, fragmentation on a length scale of 100~nm perpendicular to
this plane is obvious. This experimental observation is a striking confirmation of our original
conclusion that the high-symmetry points within the $(a,b)$ plane are essential for stabilizing
the domain vortex structure with a central meeting point of six domains and 32 symmetry. In the
$(a,c)$ or $(b,c)$ plane a point with 32 symmetry or higher does not exist which promotes the
observed fragmentation of the vortex. This result is partially supported by transmission electron
microscopy. According to Yu et al.\ domains in the plane parallel to the polarization avoid one
another so that there is no actual point in which all the domain states meet.\cite{Yu13} Zhang et
al.\ observe that the domains approach one another down to a distance of a few unit cells but
below that the distinction between domain states is no longer possible. Investigations of domain
vortices by transmission electron microscopy in the plane perpendicular to the polarization have
not been reported so far.

\begin{figure}
    \centering
        \includegraphics[width=\columnwidth,clip]{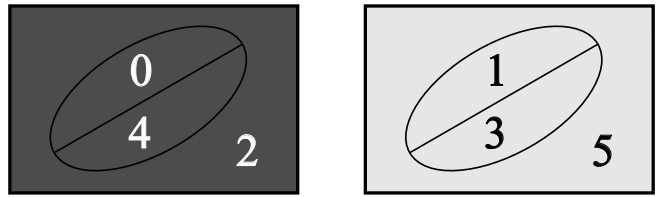}
    \caption{``Islands'' of even or odd domains which may in principle exist within a domain of the same parity.}
    \label{fig:Figure7}
\end{figure}

In summary, starting from the high-symmetry lattice points allowing the formation of a
domain-vortex pattern in YMnO$_3$ we showed that three classes of vortex-antivortex configurations
need to be considered. The connectivity rules within the domain network reveal that only one of
these classes is stabilized. In the formation of the network a number of properties regarding the
energy of the walls between different types of domain states have been concluded and found to be
in agreement with theory (with a possibility of remaining high-energy islands as shown in
Fig.~\ref{fig:Figure7}). Electron microscopy confirmed our prediction regarding the formation and
fragmentation of domain vortices of different crystallographic orientation.

\begin{figure}
    \centering
        \includegraphics[width=\columnwidth,clip]{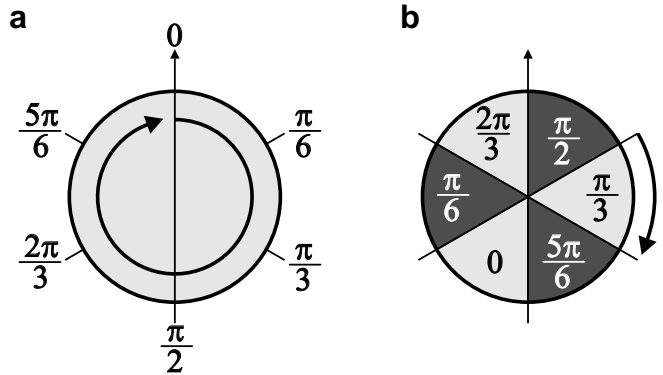}
    \caption{Angular dependence of the order-parameter phase for (a) a
topological defect and (b) a discrete domain pattern.}
    \label{fig:Figure8}
\end{figure}

\section{Other work on vortex-domains}

\subsection{Topology and scaling laws}

Griffin et al.\ interpreted the arrangement of the ferroelectric domains as the result of a
topological phase transition and the domain vortex cores as manifestation of topological defects
whose distribution is governed by universal scaling laws.\cite{Griffin12} On first glance this
seems to be incompatible with the present work. Despite a number of common properties with
topological defects the vortices in YMnO$_3$ clearly appear from our description as geometric
defects stabilized by symmetry and not for reasons of topology. Linear topological defects result
from the breaking of a continuous symmetry: gauge symmetry for superconducting and superfluid
vortices or rotational symmetry for disclinations in liquid crystals. These defects are stabilized
for purely topological reasons related to the continuous character of the order-parameter space
without any room for domains and domain walls.\cite{Mermin79} By contrast, the ferroelectric phase
in YMnO$_3$ results from a symmetry-breaking transition creating discrete domains in which domain
walls play an essential role for the formation of the domain-vortex network. More generally,
within the theory of defects the topological concept applies to the order-parameter space (in the
sense of Goldstone variables) that is the topology of the high symmetry group, and not only to
real space. It is the inter-connection between the two topologies which makes the property of
topological defects interesting, whereas in YMnO$_3$ one simply refers to real space since the
topology of the order-parameter space is discrete.

The apparent contradiction can be solved by taking into account the different temperature
dependence of the primary $K_3$ trimerization mode and the improper ferroelectric $\Gamma_2$ mode.
Right below $T_{\rm C}=1260$~K the amplitude of the polarization with respect to the trimerization
is orders of magnitude smaller than at room temperature.\cite{Fennie05} In this case the YMnO$_3$
lattice is, up to the sixth power of the order parameter, energetically isotropic with respect to
the azimuthal direction of the $K_3$ tilt mode and, hence, isotropic with respect to the phase of
the order parameter. Thus, right below $T_{\rm C}$ the direction of the $K_3$-related tilt, and
with it the phase of the of order parameter, is expected to vary continuously as sketched in
Fig.~\ref{fig:Figure8}(a). Therefore, in spite of the discreteness of the YMnO$_3$ lattice, the
phase transition at $T_{\rm C}$ would share basic properties with transitions with continuous
symmetry breaking being described by an order parameter of $U(1)$ symmetry, which allows the
formation of topological defects.

Upon cooling the ferroelectric polarization grows until it is no longer negligible. As a result,
six energetically preferred directions for the $K_3$-related tilt emerge and the effective
continuous symmetry of the order parameter breaks down into six discrete domain states. The domain
pattern sketched in Fig.~\ref{fig:Figure8}(b) is formed, and at room temperature we find the
geometrically protected network of domains that is analyzed in the present work. Ideally, the
verification of the model of continuous symmetry-breaking requires the observation of vortices
without associated walls near $T_{\rm C}$ and the progressive formation of walls upon cooling.
However, since the application of techniques like piezoresponse force microscopy in the range of
$10^3$~K is difficult, and because the temperature range in which the symmetry is effectively
continuous can be very narrow, an indirect verification with annealing cycles in the vicinity of
$T_{\rm C}$ with subsequent probing of the domain structure at ambient conditions may suffice.

\subsection{Other types of vortex domains}

Similar domain patterns as the one discussed by us have been previously discussed in different
systems. The skyrmion-vortex-like domain patterns predicted to occur in multi-axial ferroelectrics
depend only partly on the order-parameter symmetry and require finite size effects for producing
the energy-consuming depolarizing fields.\cite{Catalan12, Gruverman08} The charge-density-wave
domains in compounds like 2H-TaSe\cite{Chen81, Chen82} present six-branch domain configurations.
They represent three spatial in-plane orientations and $+$ or $-$ out-of-plane ferroelastic
strain, which, however, form an essentially different network with a variety of fragmented
configurations related to the random array of dislocations and discommensurations required for
their stabilization.\cite{Walker82, Janovec83}

The hexagonal manganites  represent, to our knowledge, the first system in which ferroelectric
domains displaying an exclusive type of vortex-like configuration have been observed. This is
because such configurations result not only from the symmetry of the order-parameter, but also
from the high-symmetry of the lattice points on which they form.

\acknowledgements

The authors thank Nicola A. Spaldin for constructive discussions and Dennis Meier for sharing the
data in Fig.~\ref{fig:Figure6} with us prior to their publication. P. T. Thanks the ETH Zurich for
supporting his stay as a guest professor. M. L. acknowledges support of his position by an ETH
Research Grant.


\end{document}